\documentclass[epj,final]{svjour}
\usepackage{graphicx}
\usepackage{amsmath}
\usepackage{mathrsfs}
\begin{document}
\title{Microscopic analysis of dipole electric and magnetic strengths in $^{156}$Gd}
\author{V.O. Nesterenko\inst{1,2}, P.I. Vishnevskiy\inst{1,2},
 P.-G. Reinhard\inst{3}, A. Repko\inst{4} and J. Kvasil\inst{5}}
\institute{Laboratory of Theoretical Physics,
Joint Institute for Nuclear Research, Dubna, Moscow region, 141980, Russia
\and State University "Dubna", Dubna, Moscow Region, 141980, Russia
\and Institut f\"ur Theoretische Physik II, Universit\"at Erlangen, D-91058, Erlangen, Germany
\and Institute of Physics, Slovak Academy of Sciences, 84511, Bratislava, Slovakia
\and Institute of Particle and Nuclear Physics, Charles University, CZ-18000, Praha 8, Czech Republic}

\date{\today}

\abstract{
The dipole electric ($E1$) and magnetic ($M1$) strengths in strongly deformed $^{156}$Gd
are investigated within a fully self-consistent Quasiparticle Random Phase
Approximation (QRPA) with Skyrme forces SVbas, SLy6 and SG2. We inspect, on the same theoretical footing,
low-lying dipole states and the isovector giant dipole resonance in $E1$ channel
and the orbital scissors resonance as well as the spin-flip giant resonance (SFGR) in $M1$ channel.
Besides, $E1$ toroidal mode and low-energy spin-flip $M1$ excitations are considered.
The deformation splitting and dipole-octupole coupling of electric excitations are analyzed.
The origin of SFGR gross structure, impact of the residual interaction and interference
of orbital and spin contributions to SFGR
are discussed. The effect of the central exchange $\textbf{J}^2$-term from the Skyrme functional
is demonstrated. The calculations show a satisfactory
agreement with available experimental
data, except for the recent NRF measurements of M. Tamkas et al for $M1$ strength at 4-6 MeV,
where, in contradiction with our calculations and previous $(p,p')$ data,
almost no $M1$ strength was observed.}

\PACS{21.60.Jz, 27.70.+q, 13.40.-f, 21.10.-k }
\maketitle

\section{Introduction}
\label{intro}
Electric and magnetic dipole excitations represent an important part
of nuclear dynamics \cite{Har01,Paar07,Hei10,Savran13,Lan19}. These excitations include
at least four nuclear modes:  i) a group of low-energy $E1$ states often called as a pygmy
dipole resonance (PDR),  which are of great importance in astrophysical reaction chains
  \cite{Paar07,Savran13,Lan19}, ii) isovector
E1 giant dipole resonance (GDR) as a benchmark for the isovector channel in modern
density functionals \cite{Paar07,Lan19,SVbas,Kleinig08}, iii)
isovector $M1$ low-energy orbital scissors
resonance (OSR) \cite{Iud78,Bohle84} as a remarkable example of a magnetic orbital flow
\cite{Iud78} and mixed symmetry states \cite{Piet_MSS}, and iv)  $M1$ spin-flip giant
resonance (SFGR) which is important to test spin-orbit splitting and tensor forces
(see, e.g.
\cite{CS_PRC09,Ves_PRC09,Nes_JPG10,Gor_PRC16,Tse_PRC19,Paar_PRC20,Mer_PRC23}).
Besides, the low-energy dipole spectrum can incorporate a toroidal $E1$ resonance
(see \cite{Paar07,Nest_PAN16} for reviews  and
\cite{Kvasil_PRC11,Rep_PRC13,Rep_EPJA17,Rep_EPJA19,Ne_PRC19}
for recent studies) and so called $M1$ "spin scissors" states (see
\cite{Bal_PRC18,Bal_PAN20} for macroscopic predictions
and \cite{Nest_PRC_SSR,Nest_PAN_SSR} for microscopic analysis).

The deformed $^{156}$Gd is one of the most suitable nuclei to investigate all these
dipole modes together, including OSR which exists only in deformed nuclei.
A large quadrupole deformation of $^{156}$Gd favors an  appearance of $E1(\Delta K=1)$
toroidal mode \cite{Rep_EPJA17} and low-energy  spin-flip excitations
\cite{Nest_PRC_SSR,Nest_PAN_SSR} (microscopic realization of the predicted "spin-scissors"
mode). What is important, for $^{156}$Gd there are experimental data
 for most of the dipole resonances listed above: GDR \cite{Gur_exp81},
OSR \cite{Bohle84,Pitz_exp89,Richter90,Richter95} and SFGR
\cite{Richter90,Richter95,Wor_exp94}. Moreover, quite recently
 the {\it separate} $E1$ and $M1$ strengths for individual
 states  at 3.1-6.2 MeV in $^{156}$Gd were measured in one and the same
NRF experiment \cite{Tamkas_NPA19}.  Thus, for the first time,
the OSR and partly PDR energy regions in a heavy strongly deformed nucleus were
experimentally explored in detail.

Dipole modes in $^{156}$Gd were already theoretically explored
in the quasiparticle random phase approximation (QRPA) \cite{Nojarov97,Sarr96,Guliev20}
and Quasiparticle-Phonon Nuclear Model (QPNM) \cite{Sol_NPA96,Sol_PPN20}.
The early studies  \cite{Nojarov97} and \cite{Sarr96} were devoted to
OSR and SFGR, respectively. A recent comprehensive QRPA analysis of $E1$ excitations (GDR and PDR)
in Gd isotopes, including $^{156}$Gd, can be found in Ref. \cite{Guliev20}.
In QPNM, low energy $E1$ and $M1$ excitation in $^{156}$Gd were scrutinized
taking into account the coupling with complex configurations \cite{Sol_NPA96,Sol_PPN20}.
All these studies were performed within {\it not self-consistent } models employing
a {\it separable} residual interaction. Further, M1 strength in $^{156-158}$Gd
was recently investigated \cite{Sasaki23} in the framework of the self-consistent noniterative finite-amplitude
method (FAM) with Skyrme force SLy4 \cite{SLy6}. The importance of the orbital and
spin M1 strengths for estimation of $(n,\gamma)$ cross sections (relevant for
the rapid neutron capture process) was demonstrated.

In this paper, we analyze $E1$ and $M1$ excitations in $^{156}$Gd within {\it
fully self-consistent} deformed QRPA  with Skyrme forces
\cite{Rep_EPJA17,Rep_arxiv,Rep_PhD,Rep_sePRC19,Kva_seEPJA19}.  Our approach was
already successfully applied to describe $E1$ \cite{Kleinig08,Rep_EPJA17,Don20} and $M1$
\cite{Ves_PRC09,Nes_JPG10,Nest_PRC_SSR,Nest_PAN_SSR,Pai16} modes in various deformed nuclei.
As compared with previous studies for $^{156}$Gd \cite{Nojarov97,Sarr96,Guliev20,Sol_NPA96,Sol_PPN20,Sasaki23},
we also inspect the deformation-induced dipole-octupole coupling of electric excitations,
spin-orbit interference in M1 states, proton-neutron gross-structure of SFGR and impact of
central exchange $\textbf{J}^2$ contribution. Furthermore, we consider toroidal and compression
$E1$ modes and discuss a possible existence of low-energy $M1$ spin-scissors resonance.

A simultaneous exploration of electric and magnetic modes within  a self-consistent
Skyrme approach based on the family of SV parametrizations \cite{SVbas}
 was recently carried out for $^{208}$Pb \cite{208Pb}. It was shown
that the chosen Skyrme forces can well describe electric modes but lead
to significant variations of the results for magnetic excitations. It was concluded
that further developments of the Skyrme functional in the spin channel are necessary.
Here we suggest a simultaneous microscopic exploration of $E1$ and $M1$
in a strongly deformed nucleus $^{156}$Gd. Such a thorough exploration, being
valuable itself, can also help to outline possible ways for a further improvement of
 Skyrme energy-density functionals.

The paper is organized as follows. In Sec. 2, the calculation scheme is outlined.
In Sec. 3, results of the calculations for
GDR, SFGR, PDR and OSR in $^{156}$Gd are considered. The low-energy dipole spectra are
compared with recent NRF data \cite{Tamkas_NPA19}.
The toroidal $E1$ and low-energy spin-flip $M1$ excitations are briefly inspected.
In Sec. 4, the conclusions are done. In Appendix A, details of the applied Skyrme functional
are outlined.

\section{Calculation scheme}

The calculations are performed within QRPA model
\cite{Rep_EPJA17,Rep_arxiv,Rep_PhD,Rep_sePRC19,Kva_seEPJA19}
 based on Skyrme functional \cite{Ben_RMP03,Stone_PPNP07}, see expression
 for the functional in Appendix A. The model is fully self-consistent since
 i) both mean field and residual interaction are derived from the same Skyrme functional,
 ii) the contributions of all time-even densities and time-odd currents from the
 functional are taken  into account, iii) both particle-hole and pairing-induced
 particle-particle channels are included, iv) the Coulomb (direct and exchange) parts
 are involved in both mean field and residual interaction. The QRPA is implemented in its matrix
 form \cite{Rep_arxiv,Rep_PhD}. Spurious admixtures caused  by violation of the translational
 and rotational invariance are removed using the technique \cite{Kva_seEPJA19}.

A representative set of Skyrme forces is used. We employ the force SLy6 \cite{SLy6} which
was shown optimal for description of $E1$ excitations in the given
framework \cite{Kleinig08}, the recently developed  force SVbas \cite{SVbas}, and the
force SG2 \cite{SG2} which is often used in analysis of $M1$ modes, see e.g.
\cite{Ves_PRC09,Nes_JPG10,Nest_PRC_SSR,Nest_PAN_SSR,Sarr96}.
As seen from Table \ref{tab-1}, these forces differ by:
i) isovector (IV) sum-rule enhancement parameter $\kappa$ which is important  for description
of $E1$ excitations \cite{Ben_RMP03,Nest08},
ii) isoscalar (IS) effective mass $m_0/m$ which can significantly influence
the single-particle spectra,
iii)  IS and IV  spin-orbit parameters  $b_4$ and  $b_4'$
(defined in Appendix A and Refs. \cite{Ves_PRC09,Stone_PPNP07}) which are crucial
for description of  $M1$ spin-flip  mode and its proton-neutron splitting
\cite{Ves_PRC09,Nes_JPG10,Nest_PRC_SSR,Nest_PAN_SSR,Sarr96},
iv) spin-dependent Landau-Migdal parameters $g_0$ and $g_0'$
characterizing IS and IV residual interaction in the spin-isospin channel
\cite{Mig67,Lan80}. Note that Skyrme and Migdal forces have some similarities
(e.g. both of them use a contact interaction) and so use of Landau-Migdal parameters
for the analysis of numerical results obtained with Skyrme functional
is relevant \cite{208Pb,SG2}. The values  of $g_0$ and $g_0'$ in Table
\ref{tab-1} are obtained from parameters of the Skyrme forces
using a bare mass normalization \cite{208Pb}.
Note also that the applied Skyrme forces use different sorts of pairing
(density-dependent surface  for SVbas and volume  for SLy6 and SG2), see details below.

\begin{table} 
\centering
\caption{IV enhancement factors $\kappa$, IS effective masses  $m_0/m$,
 IS and IV spin-orbit parameters $b_4$ and $b'_4$, IS and IV spin-dependent Landau-Migdal
 parameters $g_0$ and $g_0'$ for Skyrme forces SVbas, SLy6, and SG2.}
\label{tab-1}
\begin{tabular}{|c|c|c|c|c|c|c|}
\hline
 force & $\kappa$ & $m_0/m$ & $b_4$ & $b^{'}_4$ & $g_0$ & $g_0'$  \\
       &          &         & MeV $\rm{fm}^5$ &  MeV $\rm{fm}^5$ & & \\
 \hline
SVbas & 0.4 &  0.90 & 62.3 & 34.1 & 0 & 1.1 \\
SLy6 & 0.25 & 0.69 &  61.0 & 61.0 & 2.0 & 1.3 \\
SG2 &  0.53 & 0.79 & 52.5 & 52.5  & 0 & 0.6 \\
\hline
\end{tabular}
\\
\end{table}
\begin{table} 
\centering
\caption{Deformation parameters $\beta_2$, proton and neutron pairing gaps $\Delta_p$
and $\Delta_n$ and energies of $2^+_1$ state of the ground-state rotational band,
calculated with Skyrme forces  SVbas, SLy6 and SG2. The experimental data
are from database \cite{nndc}.
}
\label{tab-2}
\begin{tabular}{|c|c|c|c|c|c|c|}
\hline
       & $\beta_2$  & $\Delta_p$ & $\Delta_n$ &  $E_{2^{+}_1}$\\
       &            &  MeV       &   MeV      & keV \\
\hline
 exper & 0.34       &            &           & 89   \\
\hline
SVbas &  0.327  &  0.83  & 0.98 &  106 \\
\hline
SLy6 & 0.333  &   0.77 & 0.43 &  63 \\
\hline
 SG2 (no $\textbf{J}^2$) &  0.329   & 0.85 & 0.81 & 90 \\
\hline
 SG2 (with $\textbf{J}^2$) &  0.318   & 0.87 & 0.96 & 103 \\
\hline
\end{tabular}
\\
\end{table}

It is known that the so-called tensor $\textbf{J}^2$-term in the Skyrme functional
(see Appendix \ref{Sec:_Skyrme}) can affect spin-orbit splitting
\cite{CS_PLB07,Les_PRC07,Bender_PRC09,Satula_IJMPE09} and M1 modes \cite{Ves_PRC09,Nes_JPG10}.
Skyrme forces from our set (SVbas, SLy6 and SG2) were fitted without taking into account this term
\cite{SVbas,SLy6,SG2}.
However it is worth to try to estimate its impact at least by a perturbative way.
We chose for this aim the force  SG2 which was specially fitted for the spin-isospin channel.
 All SG2 calculation in this study are performed with $\textbf{J}^2$-term (with exception of some
 illustrative cases used  for the comparison). In general, $\textbf{J}^2$-term is produced
 by central exchange and non-central tensor interactions \cite{CS_PLB07,Les_PRC07,Bender_PRC09}.
 For simplicity, we limit ourselves by the central exchange contribution to $\textbf{J}^2$.
 To our knowledge, even such limited case was not yet actually  analyzed for the SFGR in deformed nuclei.

The mean field spectra and pairing characteristics are calculated by the code
SKYAX \cite{SKYAX} using a two-dimensional grid in cylindrical coordinates.
The calculation box extends up to three nuclear radii, the grid step
is 0.4 fm. The axial quadrupole equilibrium deformation is obtained by minimization
of the energy of the system. As seen from Table~\ref{tab-2}, deformation parameters
$\beta_2$ for SVbas, SLy6 and SG2 (no $\textbf{J}^2$) are close to the experimental value.
The $\textbf{J}^2$-contribution somewhat decreases $\beta_2$ for SG2.

The pairing is described by the zero-range pairing interaction \cite{Be00}
\begin{equation}
  V^{q}_{\rm pair}(\textbf{r},\textbf{r}') =  G_{q}
\Big[ 1 - \eta \: \Big(
\frac{\rho(\textbf{r})}{\rho_{\rm pair}}
\Big)\Big]
\delta(\textbf{r}-\textbf{r}')
\label{Vpair}
\end{equation}
where $G_{q}$ are proton ($q=p$) and neutron ($q=n$) pairing strength constants
fitted to reproduce empirical pairing gaps along selected
isotopic and isotonic chains \cite{G_Rein}.
Further, $\rho(\textbf{r})=\rho_p(\textbf{r})+\rho_n(\textbf{r})$ is the sum of
proton and neutron densities. We switch on the volume pairing with
$\eta$=0 (SLy6, SG2)  and the density-dependent surface pairing with $\eta$=1
(SVbas). The model parameter $\rho_{\rm pair}$=0.2011 ${\rm fm}^{-3}$ for surface pairing
is  determined in the fit for SVbas \cite{SVbas}.
Pairing is calculated within HF-BCS
(Hartree-Fock and Bardeen-Cooper-Schrieffer) method \cite{Rep_EPJA17,SKYAX}.
To cope with a divergent character of zero-range pairing forces,
an energy-dependent cut-off is used \cite{Rep_EPJA17,Be00}. The calculated
proton $\Delta_p$ and neutron $\Delta_n$ pairing gaps are shown in Table~\ref{tab-2}.

The QRPA calculations use a large configuration space. For example, the
single-particle basis for SVbas includes 683 proton and 787 neutron levels.
For $E1$ excitations, the energy-weighted sum rule
\begin{equation}
\label{eq:EWSR}
 EWSR=\frac{\hbar^2e^2}{8\pi m}9\frac{NZ}{A}(1+\kappa)
\end{equation}
is exhausted by 99\% (SLy6) and 100\% (SVbas, SG2).

The moments of inertia $J$ are calculated in the framework of  Thouless-Valatin
model \cite{TV62} using the QRPA spectrum \cite{Rep_sePRC19}. The energy
 of the  first state in the ground-state rotational band is estimated as $E_{2^+_1}=3 \hbar^2/J$.
 This energy is sensitive to both deformation and pairing. As seen in Table~\ref{tab-2},
SVbas and SG2 (with $\textbf{J}^2$) overestimate and SLy6 underestimates this energy while SG2
(no $\textbf{J}^2$) gives a nice agreement.

The reduced  probability for electric dipole transition $0^+0 \to 1^-K_{\nu}$ ($K$=0,1)
between the ground state  $0^+0$ and $\nu$-th QRPA state $1^-K_{\nu}$ reads
\begin{equation}
\label{BE1}
B_{\nu}(E1,K)=(1+\delta_{K,1})|\:\langle\nu|\:\hat{\Gamma}(E1,K)\:|0\rangle \:|^2
\end{equation}
with
\begin{equation} \label{eq:E1}
 {\hat \Gamma}(E1,K) =
  e \sum_{q \epsilon p,n} e_{\text{eff}}^q
\sum_{k \epsilon q} [r Y_{1K}(\Omega)]_k
\end{equation}
where $Y_{1K}$ is the spherical harmonic, $e_{\text{eff}}^p=N/A$
and $e_{\text{eff}}^n=-Z/A$ are the effective charges.

For $M1$ transitions, we have
\begin{equation}
\label{BM1}
  B_{\nu}(M1,K)=(1+\delta_{K,1})|\:\langle\nu|\:\hat{\Gamma}(M1,K)\:|0\rangle \:|^2
\end{equation}
with
\begin{equation} \label{eq:M1}
 {\hat \Gamma}(M1,K) =
 \mu_N \sqrt{\frac{3}{4\pi}}\sum_{q \epsilon p,n}\sum_{k \epsilon q}
[g^{q}_s {\hat s}_k(K) + g^{q}_l {\hat l}_k (K)] .
\end{equation}
Here $\mu_{N}$ is the nuclear magneton; $g^{q}_s$ and  $g^{q}_l$ are spin and
orbital gyromagnetic factors; ${\hat s}_k(K)$ and ${\hat l}_k(K) $ are $K$-components
of the spin and orbital operators. In the present calculations, we use
 $g^{q}_s=q\widetilde{g}^{q}_s$, where
$\widetilde{g}^{p}_s=5.56$ and $\widetilde{g}^{n}_s=-3.83$ are bare
 g-factors and $q=0.7$ is the quenching parameter. The orbital g-factors
 are $g^{p}_l=1$, $g^{n}_l=0$. For calculation of SFGR, where only spin contribution is relevant
 \cite{Ves_PRC09,Nes_JPG10}, we use $g^{p}_l=g^{n}_l=0$.

 Reduced transition probabilities (\ref{BE1}) and (\ref{BM1}) are used for
 calculation of the strength functions
 \begin{eqnarray}
 \label{SFE1}
 S(E1,K;E) &=& \sum_{\nu} E_{\nu} B_{\nu}(E1,K)\zeta(E_{\nu}, E),\\
 \label{SFM1}
 S(M1,K;E) &=& \sum_{\nu} B_{\nu}(M1,K)\zeta(E_{\nu}, E)
 \end{eqnarray}
 where
$  \zeta(E - E_{\nu}) = \Delta /[2\pi[(E- E_{\nu})^2+\frac{\Delta^2}{4}]]$
is a Lorentz weight with an averaging parameter $\Delta$. The $E1$ photoabsorption
cross section reads
\begin{equation}
 \sigma (E)= \alpha \sum_{K=0,1} S(E1,K;E) ,
\end{equation}
where $\alpha = 16\pi^3/(137\cdot 9 e^2)= 0.40/e^2$. The Lorentz weight is used
for the convenience of comparison of the calculated strength with experimental data.
It simulates smoothing effects beyond QRPA: escape width and coupling to complex
configurations (CCC).  The higher the excitation energy $E$, the larger the density
of states. So one may expect
an increase of CCC-produced width with $E$. Since the $E1$ GDR lies
at much higher excitation energy (10-20 MeV) than the $M1$ SFGR (5-10 MeV), it is reasonable to use
for GDR a larger smoothing $\Delta$ than for SFGR. Following our
previous calculations  for GDR \cite{Kleinig08} and  SFGR \cite{Ves_PRC09,Nes_JPG10},
we use here  $\Delta$=2 MeV for GDR and  $\Delta$=1 MeV for SFGR.

Further  observables are the toroidal and compression isoscalar strengths, $B_{\rm tor}(E1K,IS)$
and $B_{\rm com}(E1K,IS)$, which are calculated  using current-dependent vortical
and compression transition operators  from Refs. \cite{Rep_PRC13,Rep_EPJA17,Rep_EPJA19}.

In deformed nuclei, the dipole-octupole coupling takes place. So, unlike previous schematic
QRPA studies for $^{156}$Gd \cite{Guliev20}, our self-consistent
calculations for dipole electric states include both dipole and octupole residual interactions.

\begin{figure*} 
\centering
\includegraphics[width=13.5cm,angle=0]{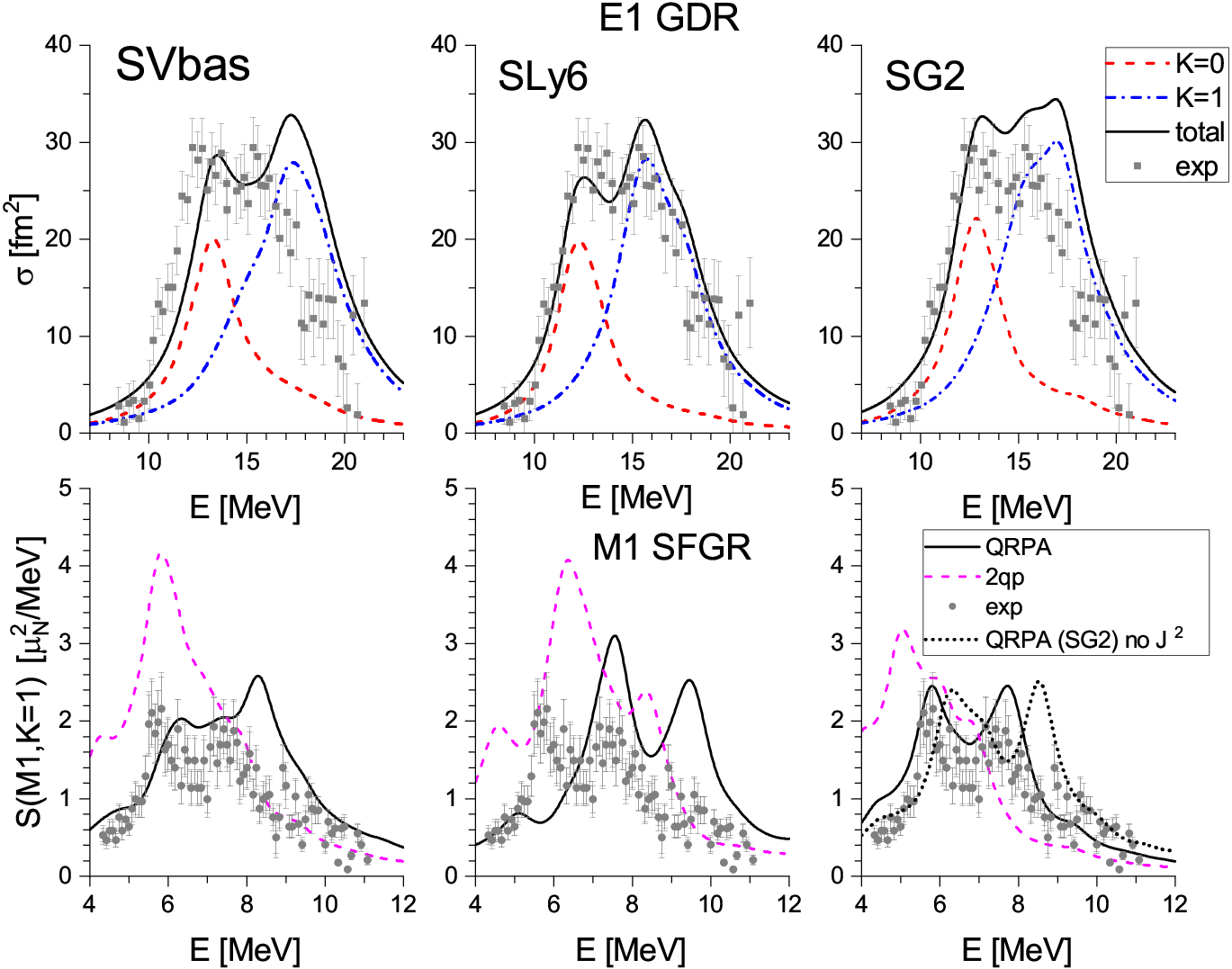}
\vspace{0.1cm}
\caption{QRPA $E1$ photoabsorption (upper panels) and $M1$ SFGR (bottom panels) strengths
in $^{156}$Gd, calculated with the forces SVbas, SLy6, and SG2. For $E1$ GDR, the
 $K=0$ (left-side red dashed line) and  $K=1$ (right-side blue dash-dot line) strengths
  are also depicted. For $M1$ SFGR, the unperturbed 2qp strength (magenta dashed line)
  is shown. For SG2, QRPA M1 strength is exhibited with (solid line) and without (dotted line)
 $\textbf{J}^2$ contribution. $E1$ GDR and $M1$ SFGR strengths use Lorentz weights with
 $\Delta$=2 MeV and 1 MeV, respectively. The strengths are compared with the
 photoabsorption data for $E1$ GDR \cite{Gur_exp81}
and $(p,p')$ data for $M1$ SFGR \cite{Wor_exp94}.}
\label{fig1}
\end{figure*}

\section{Results and discussion}
\subsection{$E1$ and $M1$ giant resonances}

As a first step, we apply SVbas, SLy6  and  SG2 to describe the
IV $E1$ GDR and $M1$ SFGR. In Fig.~\ref{fig1}, the calculated strength functions
 (\ref{SFE1}) and (\ref{SFM1}) are compared
with the photoabsorption data for $E1$ GDR \cite{Gur_exp81} and $(p,p')$
data for $M1$ SFGR \cite{Wor_exp94}. As discussed above,
E1 and $M1$ strengths use different Lorentz averaging parameters:
$\Delta$=2 MeV and 1 MeV, respectively. The $(p,p')$ data
\cite{Wor_exp94} are given in arbitrary units. So, in this case, not
the absolute scale but the distribution of strength is of interest.
The $(p,p')$ reaction excites spin-flip and not orbital
M1 strength \cite{Har01}. So the calculations for SFGR
are performed with  $g^{q}_l$=0.
\begin{figure} 
\centering
\vspace{0.1cm}
\includegraphics[width=8.5cm,angle=0]{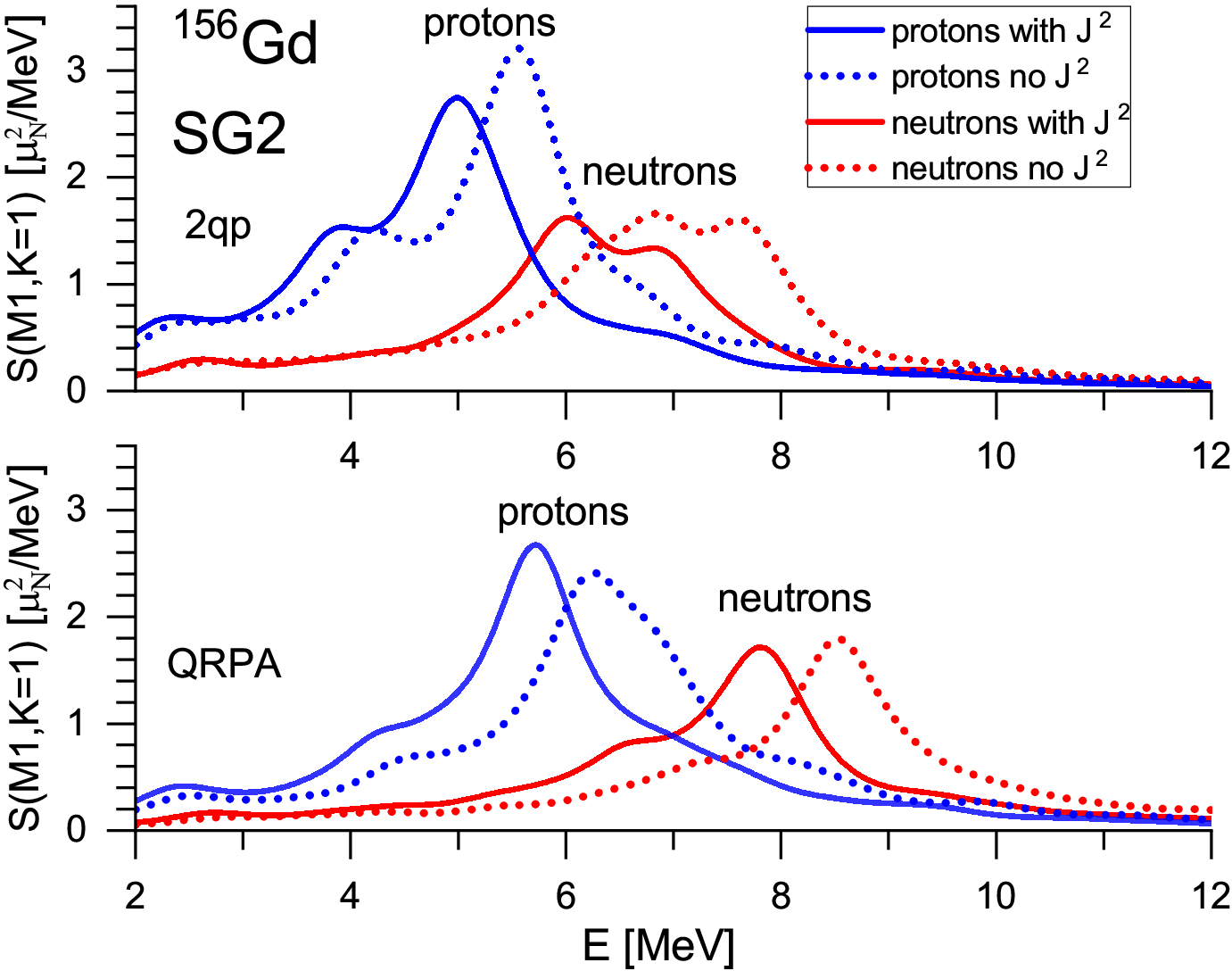}
\caption{SG2 2qp (upper panel) and QRPA (bottom panel) proton and neutron spin M1 strength
functions in $^{156}$Gd, calculated with and without $\textbf{J}^2$-impact.}
\label{fig2_np_spin_M1}
\end{figure}

In the upper panels of Fig.~\ref{fig1}, we present GDR branches $K=0$ and $K=1$
and the total strength.
In general, all three Skyrme forces (with some preference for SLy6)
give a nice description of GDR, including its
deformation splitting into $K=0$ and $K=1$ branches.
A small overestimation of the peak height of the $K=1$ branch can be explained by
insufficient smoothing. The $K=1$ branch is located at a higher energy than $K=0$
and so should acquire more broadening from CCC.
This could be simulated by an energy dependent folding width which we
avoid here to keep the analysis simple.  SG2 calculations for GDR are
performed with $\textbf{J}^2$-term. As we checked, the impact of this term on GDR is negligible.

The bottom panels of Fig.~\ref{fig1} show the $K=1$ branch of $M1$
strength, responsible for the SFGR \cite{Har01,Hei10,Ves_PRC09,Nes_JPG10}, and $(p,p')$
experimental data \cite{Wor_exp94}. Note that, in the recent FAM study \cite{Sasaki23},
the calculated and experimental  distributions of
M1 strength were not compared. Fig.~\ref{fig1} demonstrates that,
in accordance to experiment \cite{Wor_exp94}, our calculated M1 strengths exhibit two large peaks
produced by proton and neutron spin-orbit excitations. The origin of
these peaks is illustrated in Fig.~\ref{fig2_np_spin_M1} where the separate
proton and neutron spin ($g^{p}_l=g^{n}_l=0$) M1 QRPA strengths are depicted
for 2qp and QRPA cases. It is known that unperturbed 2qp spin-flip proton  and neutron
excitations are determined by spin-orbit interaction
$V_{so} \propto (\textbf{l} \cdot \textbf{s})$ where $\textbf{l}$ and  $\textbf{s}$
are orbital moment and spin of the nucleon. In $^{156}$Gd, the number of
neutrons $N$  is much larger than number of protons $Z$. So, as compared with protons,
neutrons occupy a higher valence quantum shell with a  larger orbital moments $l$.
This  leads to a larger neutron spin-orbit splitting and so to a higher neutron
spin-flip energy.  Fig.~\ref{fig2_np_spin_M1} shows that indeed the neutron spin-flip
M1 mode has a higher energy than the proton one. The proton peak is higher because
$|g^{p}_s| > |q^{n}_s|$. The residual interaction  upshifts  M1 strength but still keeps the
two-peak proton-neutron gross structure. A detail discussion of the proton-neutron gross
structure of SFGR can be found elsewhere \cite{Hei10,Ves_PRC09,Nes_JPG10}.
\begin{figure*} 
\centering
\includegraphics[width=15cm,angle=0]{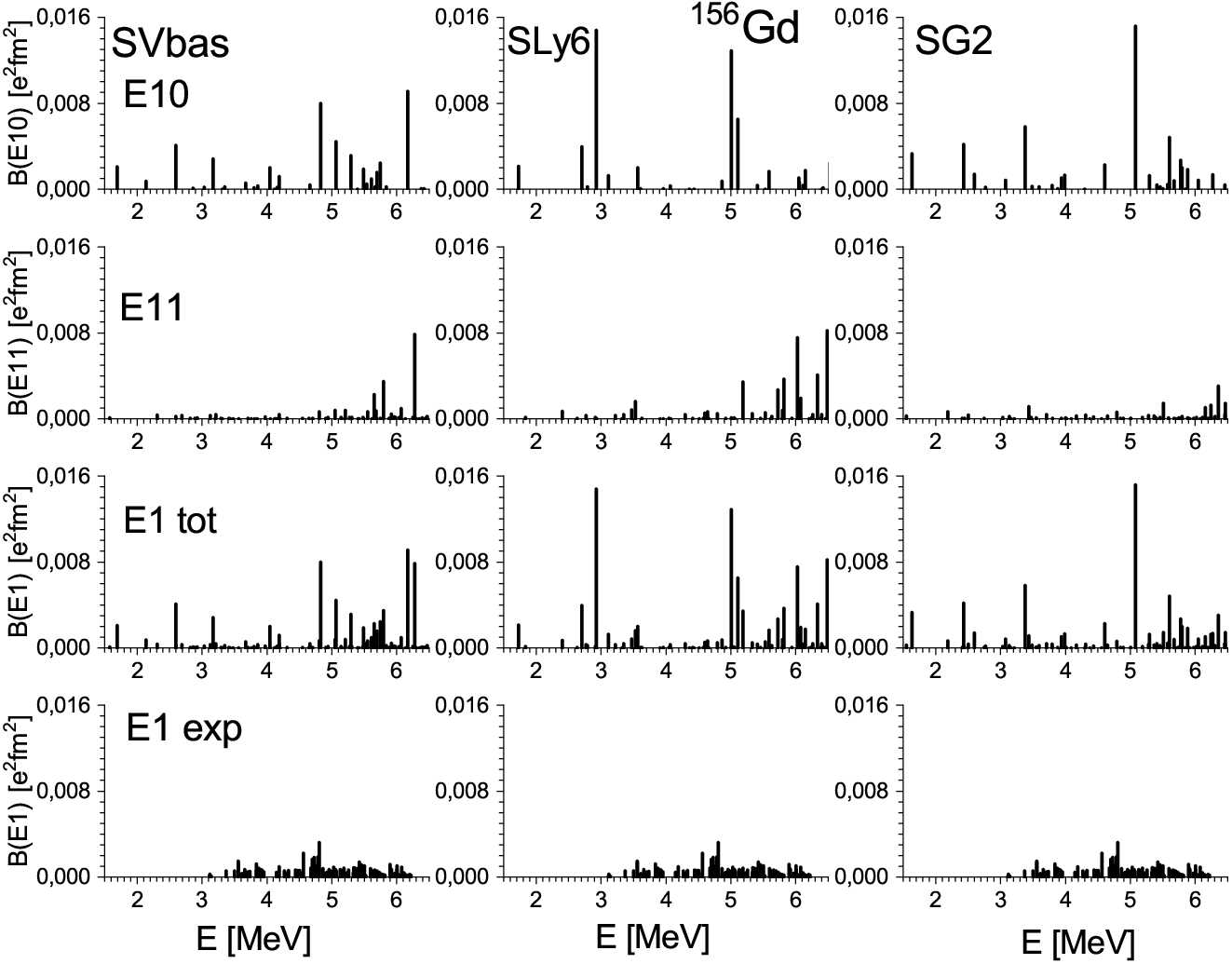}
\caption{The calculated (SVbas, SLy6, SG2) $K=0$, $K=1$ and total low-energy $E1$ strengths in
$^{156}$Gd as compared with $E1$ experimental data \cite{Tamkas_NPA19}.}
\label{fig3_E1}
\end{figure*}

\begin{figure*} 
\centering
\includegraphics[width=16cm,angle=0]{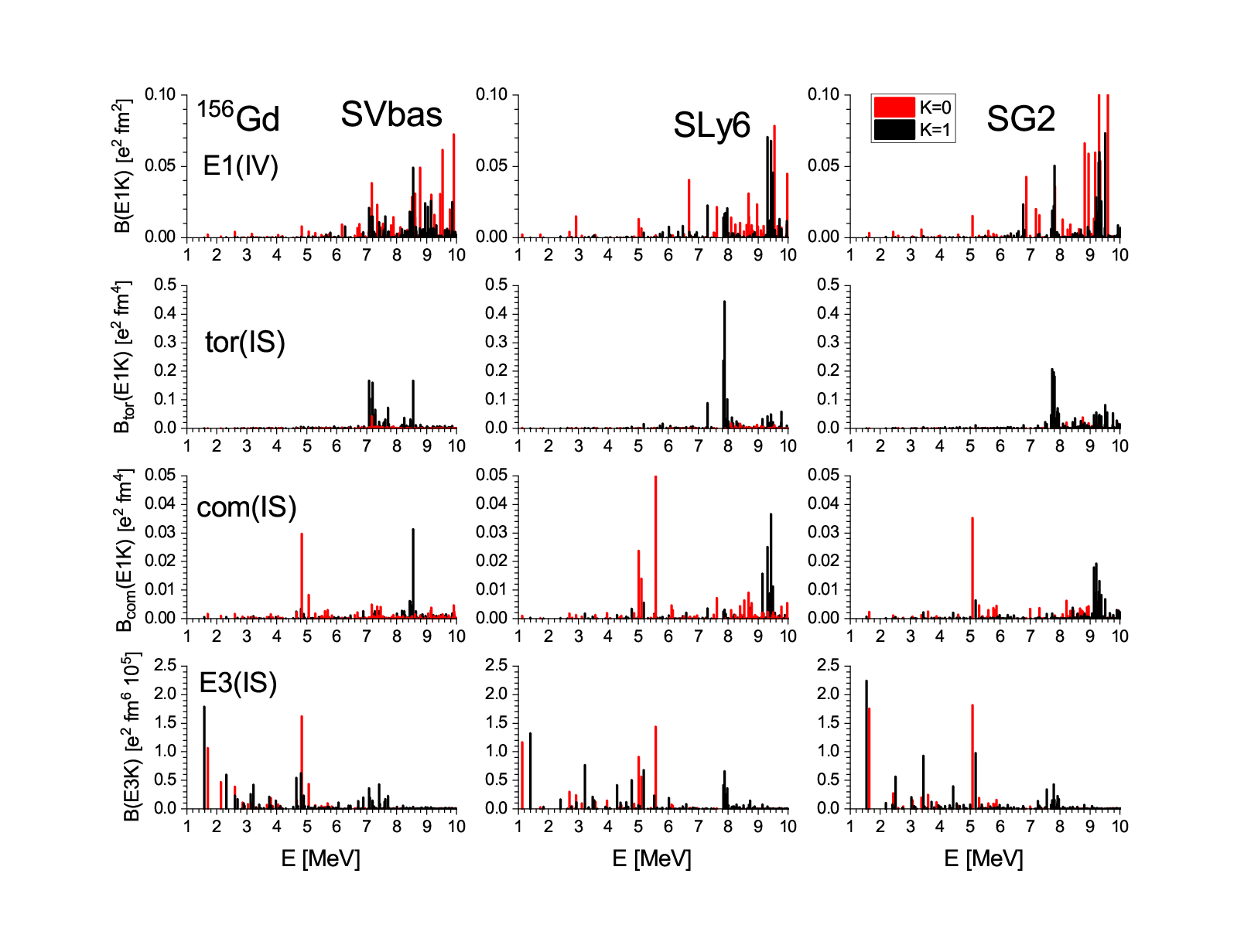}
\caption{ $K=0$ (red) and $K=1$ (black) branches of isovector $E1$ and
isoscalar toroidal,  compression and octupole
strengths in $^{156}$Gd, calculated with the forces SVbas, SLy6 and SG2.}
\label{fig4_E1tc}
\end{figure*}

As seen from bottom panels of Fig.~\ref{fig1}, all Skyrme forces well describe
the energy difference $\Delta E \approx $ 2.1 MeV between M1 peaks. At the same time,
energies of $M1$ peaks are overestimated by 0.7-0.8 MeV (SVbas), 1.8-2.0 MeV
(SLy6) and 0.7-0.9 MeV (no $\textbf{J}^2$ in SG2).  Following previous studies
\cite{Ves_PRC09,Nes_JPG10}, the peak energies are mainly determined by two
factors: single-particle spin-orbit splitting and upshift due to the
IV residual interaction in the spin channel of Skyrme
functional. The latter effect is seen by comparison of unperturbed
two-quasiparticle (2qp) and QRPA $M1$ strengths in Fig.~\ref{fig1}.
It is known that IS (IV) residual interaction leads to the energy downshift (upshift)
of the strength. In Fig.~\ref{fig1}  we see a significant upshift
of M1 spin-flip strength, which points to the dominant impact of IV residual interaction.
The impact of IV spin-spin interaction in Fig.~\ref{fig1} is obviously too strong and
CCC will hardly help to correct the SFGR energy.
However, if we include $\textbf{J}^2$-term, as done for SG2, then a necessary downshift
of M1 strength is obtained and a good agreement with the experiment is achieved.
So our results indicate a need for $\textbf{J}^2$ contribution even if this impact is limited
by central exchange contribution.

The impact of $\textbf{J}^2$-term is additionally inspected for SG2 in Fig.\ref{fig2_np_spin_M1}.
We see that this term leads to  a considerable change already in 2qp proton and neutron strengths,
i.e. at the mean-field stage. Namely, we get a strong downshift of both proton and neutron peaks.
Instead, the $\textbf{J}^2$ impact to the residual interaction is of a minor
importance. Note that these results should be considered as preliminary.
First of all, inclusion of the true tensor interaction with the corresponding refit
of the force parameters can be important \cite{Ves_PRC09,Nes_JPG10,CS_PLB07,Les_PRC07,Bender_PRC09}.
Further, the present SG2 results should be checked by calculations with other Skyrme
parametrizations, which we leave to a  future work.
Anyway, our calculations show that a thorough test of
spin-isospin parameters of the Skyrme functional by SFGR data for various nuclei is desirable.
Note that, in previous studies \cite{CS_PLB07,Les_PRC07,Bender_PRC09}, the impact of $\textbf{J}^2$-term
was mainly tested by description of single-particle spectra, in particular of spin-orbit splitting
of single-particle levels. However, single-particle spectra are sensitive to many factors and, in this sense,
are not optimal for $\textbf{J}^2$ tests. Instead, the energy and two-peak gross-structure
of SFGR are more stable characteristics and so are more suitable for the tests.

An additional analysis can be done in terms of IS and IV Landau-Migdal spin-isospin
parameters $g_0$ and $g_0'$ shown in Table \ref{tab-1}. SG2 values are obtained
with $\textbf{J}^2$, otherwise, we have $g_0$=0.8 and $g_0'$=1.2.
We see that SVbas, SLy6 and SG2
have different strengths of IS and IV  residual interactions in the spin-isospin channel.
Nevertheless, despite significant variations in the interaction,
SVbas, SLy6 and SG2 (no $\textbf{J}^2$) significantly overestimate the SFGR energy.
 The presence of IS interaction in SLy6 and SG2 (no $\textbf{J}^2$) does not improve the SFGR description.
 The latter looks reasonable since IV spin g-factor $g_s^{IV}=g_s^{p}-g_s^{n}$=6.24 is much larger
 than IS spin g-factor  $g_s^{IS}=g_s^{p}+g_s^{n}$=1.35. So IS fraction of SFGR should
 be very small even in the case of the significant IS spin-spin residual interaction. Altogether,
 SFGR is mainly governed by IV residual interaction and does not suit for
 manifestation of IS one. However, IS spin-spin residual
 interaction can sometimes lead to local effects in M1 strength function, see
 \cite{Nes_JPG10,Paar_PRC20} for details.

\begin{table}  
\centering
\caption{Calculated and experimental \cite{Tamkas_NPA19} summed $B(E1)$ strengths
(in $\rm{e^2 fm^2}$) at 3.1 - 6.2 MeV in $^{156}$Gd.}
\label{tab3}
\begin{tabular}{|c|c|c|c|}
\hline
 force & $K=0$ & $K=1$ & total  \\
\hline
SVbas & 0.040   &  0.014  &  0.054  \\
\hline
SLy6  & 0.030   & 0.028  & 0.058   \\
\hline
SG2 & 0.042  & 0.008  & 0.050  \\
\hline
exper. \cite{Tamkas_NPA19} & & & 0.073 \\
\hline
\end{tabular}
\end{table}

 \subsection{Low-energy conventional, toroidal and compression  $E1$ strengths}

Figure~\ref{fig3_E1} shows $E1$ strengths for excitations at 1.5 - 6.5 MeV,
calculated with the forces SVbas, SLy6 and SG2 (with $\textbf{J}^2$). Note that NRF
experiment \cite{Tamkas_NPA19} covers the energy region 3.1-6.2
MeV. We see that, in agreement with this experiment, all three Skyrme
forces produce many QRPA states in the range 3.1-6.2 MeV. Most of the strength
comes from the $K=0$ branch. This could be a consequence
of the deformation splitting of the low-energy E1 strength, when
$K=0$ branch is downshifted from the PDR energy region and becomes
dominant at $E < $ 6 MeV.

Following Table~\ref{tab3}, the calculated summed $E1$ strength is in acceptable agreement with data
\cite{Tamkas_NPA19}. A small underestimation of the experimental strength  can be explained by omitting the
CCC which can somewhat redistribute $E1$ strength. For
some states, our QRPA calculations give appreciably large $B(E1)$ values
which are not seen in data \cite{Tamkas_NPA19}.  This can be again
explained by missing CCC. In general, all three
applied Skyrme forces produce rather similar results.
Following Fig.~\ref{fig1}, the PDR region (5-9 MeV) in $^{156}$Gd lies safely below GDR.
Fig.~\ref{fig3_E1} demonstrates a
concentration of $E1$ strength at the low-energy part of this energy range.
 The summed low-lying $E1$ strength is given in Table~\ref{tab4}.

The E1 (IV) $K=0$ and $K=1$ branches are exhibited  at a wider energy range in upper panels of
Fig.~\ref{fig4_E1tc}. The figure also includes IS toroidal E1, compression E1 and E3 strengths.
The later is calculated with $e_{\text{eff}}^p=e_{\text{eff}}^n=$1. It is seen that all three Skyrme 
forces give similar results. Like in our previous studies for spherical \cite{Rep_PRC13,Rep_EPJA19} and deformed
\cite{Nest_PAN16,Rep_EPJA17,172Yb_E1tor,Ne_PRL18} nuclei, toroidal states lie
in the PDR energy region 7-10 MeV. Moreover, in accordance with earlier studies for deformed nuclei,
the toroidal resonance is mainly presented in $K=1$ branch.

\begin{table} 
\caption{Calculated summed $B(E1)$ (in $\rm{e^2 fm^2}$), for $K=0, 1$ and total cases at
the PDR energy region 5-9 MeV in $^{156}$Gd.}
\centering
\label{tab4}
\begin{tabular}{|c|c|c|c|}
\hline
 force & $K=0$ & $K=1$ & total \\
\hline
 SVbas &  0.34 &  0.34 & 0.68 \\
\hline
SLy6 & 0.25  & 0.18 & 0.43 \\
\hline
SG2 & 0.32  &  0.21 &  0.53 \\
\hline
\end{tabular}
\end{table}

\begin{figure} 
\centering
\includegraphics[width=8.5cm,angle=0]{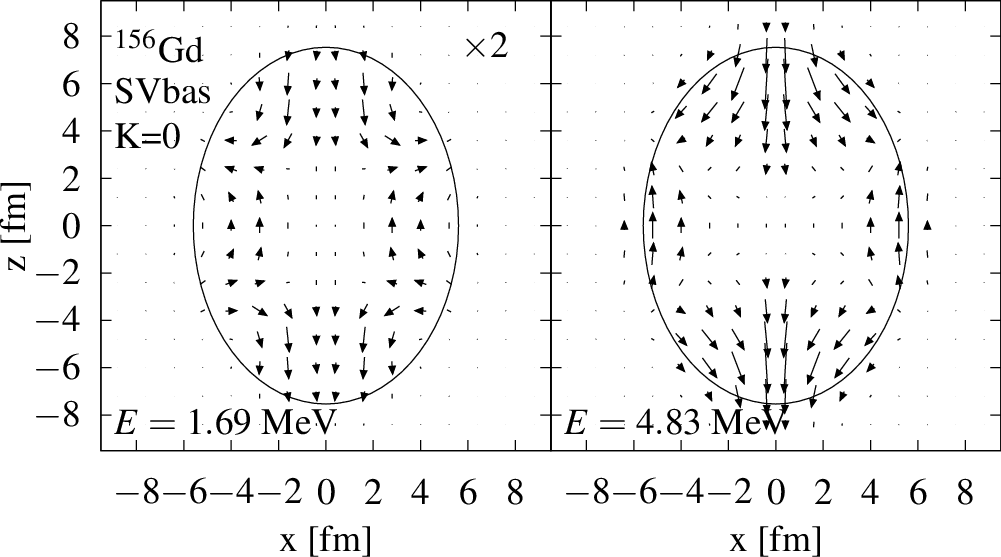}
\caption{SVbas isoscalar current transition densities for $K^{\pi}=0^-$ states at 1.69 MeV (left) and
4.8 MeV (right) in  $^{156}$Gd.  The scale of arrows in the left panel is
twice larger than in the right panel. The solid line marks the nucleus boundary.}
\label{fig5_currents}
\end{figure}

\begin{figure*} 
\centering
\includegraphics[width=15cm,angle=0]{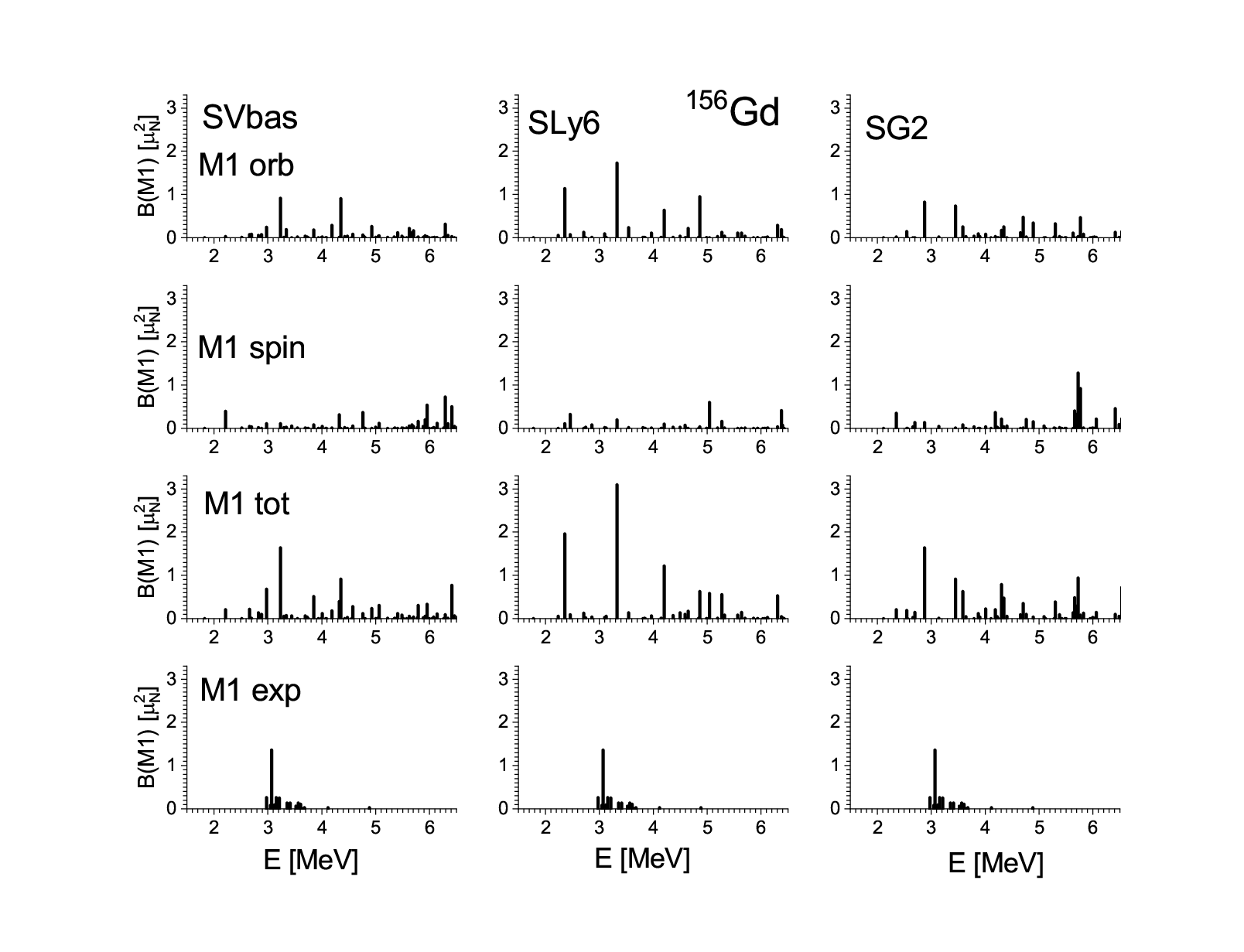}
\caption{The calculated (SVbas, SLy6, SG2) orbital, spin, and total low-energy $M1$ strengths in
$^{156}$Gd as compared with $M1$ experimental data \cite{Tamkas_NPA19}.}
\label{fig6_M1}
\end{figure*}

In Fig.~\ref{fig4_E1tc}, the compression strength is much weaker than toroidal one.
The compression mode demonstrates a clear deformation splitting typical
for electric giant resonances in prolate nuclei.
Namely, $K=0$ strength lies generally lower than $K=1$ one. Furthermore, unlike the upper panel of
Fig.~\ref{fig4_E1tc} for IV E1 strength, the compression IS E1 strength is not distributed among
many dipole states but mainly concentrated in a pair of very collective peaks,
e.g. at 4.8 MeV ($K=0$) and 8.6 MeV ($K=1$) for SVbas.  A similar pattern takes place for
 SLy6 and SG2.

 Note that energy region 4-9 MeV includes  so-called Low-Energy Octupole
 Resonance (LEOR) \cite{Har01,LEOR_exp,LEOR_MNS}. Indeed,
 the bottom panel of Fig.~\ref{fig4_E1tc} shows large IS octupole transition
 probabilities $B(E30)$ and $B(E31)$ for $I^{\pi}K=3^-0$ and $3^-1$
 states. In deformed $^{156}$Gd, the dipole and octupole modes should be mixed
 and this is clearly seen from comparison of the panels with compression dipole and
 octupole IS modes. The mixing depends on the
 structure of the state. For example, in SVbas calculations, the collective basically
 octupole $K=0$-states at 1.7 and 4.8 MeV have negligible and significant E1 compression
 fractions, respectively.

Mainly octupole character of 4.8-MeV state is confirmed by  Fig.~\ref{fig5_currents}  where
convective current  transition density (CTD) $\delta \vec{j}(x,z)$  (see the definition
in refs. \cite{Rep_EPJA17,Rep_EPJA19,Ne_PRC19}) in z-x plane for 1.7-MeV  and 4.8-MeV states are compared.
The lengths of arrows $\delta \vec{j}(x,z)$ are properly scaled for the convenience of visibility.
 Following our analysis, both 1.7-MeV  and 4.8-MeV states are  mainly IS. Their CTD are rather
 similar and well match a typical octupole flow (see for comparison Fig. 10 in ref. \cite{Se83}
 for $3^-$-state in $^{208}$Pb).
 So 4.8-MeV state is basically octupole. At the same time, its flow  somewhat reminds a pattern
  for the high-energy compression mode: the motion in pole
 regions against the quasi-static central part creates compression and decompression
 areas in the interface of moving and static parts. The similarity between E3 and compression E1
 flows is not surprising since both modes are produced by similar shell transitions and have a
 similar radial dependence $r^3$.

  The same analysis shows that SVbas $K=1$ state at 8.5 MeV  has even more complicated composition.

 The significant dipole-octupole mixture found in this study points that calculations for
 electric dipole $K=0,1$  states in deformed nuclei should take into account both dipole and octupole
 residual interactions. This was also found in the previous Skyrme QRPA studies for Nd and Sm isotopes
 \cite{Yoshida_PRC13} and highly deformed light nuclei \cite{Ne_PRC19,Ads21}.
 Note that, in schematic QRPA calculations  for dipole states in $^{156}$Gd \cite{Guliev20},
 the octupole residual interaction was omitted.

\subsection{Low-energy orbital and spin-flip $M1$ strengths}

In Fig.~\ref{fig6_M1}, the calculated orbital, spin and total (orbital + spin)
low-energy $M1(K=1)$ strengths in $^{156}$Gd are compared with NRF  $M1$ experimental
data at 2.95-6.25 MeV \cite{Tamkas_NPA19}. SG2 results are obtained taking into account
$\textbf{J}^2$-term. Note that NRF measurements should embrace
both orbital and spin-flip M1 strengths. The calculations show a wide fragmentation
of $M1$ strengths throughout the whole interval 1-7 MeV.
In the OSR energy region 3-4 MeV,  the calculated orbital strength is locally concentrated
in accordance with NRF data \cite{Tamkas_NPA19}.h e properties of some selected calculated
states are given in Table~\ref{tab-5}.

\begin{table*} 
\centering
\caption{Properties of particular QRPA low-energy $K^{\pi}=1^+$ states in $^{156}$Gd:
excitation energy $E$, orbital, spin and total $B(M1)$-values,
and main proton (pp)  and neutron (nn) components (Nillson quantum numbers
and contribution in $\%$ to the state norm).}
\label{tab-5}       
\begin{tabular}{|c|c|c|c|c|c|c|c|}
\hline
Force & E [MeV] &\multicolumn{3}{c|}{$B(M1) \; [\mu^2_N]$} & \multicolumn{2}{c|}{main 2qp components}
\\
\hline
 &    & orb & spin & total &  $[N,n_z,\Lambda]_1,[N,n_z,\Lambda]_2$  & $\%$  \\
 \hline
SVbas  & 2.21 & 0.03 & 0.39 & 0.21  & pp $[411\downarrow,411\uparrow]$ & 99\\
   &&&&&& \\
 & 3.23 & 0.92 & 0.10 & 1.63 & nn $[521\uparrow,530\uparrow]$  & 39\\
&   &  &  & & pp $[402\uparrow,411\uparrow]$ & 15\\
\hline
 SLy6  & 2.46 & 0.07 & 0.32 & 0.09 &  pp $[411\downarrow,411\uparrow]$ & 98\\
   &&&&&& \\
  &  3.33 & 1.76 & 0.20 & 3.10 &  pp $[532\uparrow,541\uparrow]$ & 41\\
         & &  &  &  & nn $[521\uparrow,530\uparrow]$ & 39\\
 \hline
 SG2   & 2.35 & 0.02 & 0.34 & 0.20 & pp $[411\downarrow,411\uparrow]$ & 99\\
   &&&&&& \\
 & 2.87  & 0.82 & 0.14 & 1.64  & pp $[532\uparrow,541\uparrow]$ & 48\\
          & &  &  &  &  nn $[642\uparrow,651\uparrow]$  & 47\\
 \hline
\end{tabular}
\\
\end{table*}

\begin{table} 
\caption{Calculated orbital, spin and total strength summed
B(M1) (in $\mu^2_N$) in OSR region 2.5-4 MeV as compared with
experimental data for $^{156}$Gd and $^{158}$Gd.}
\centering
\label{tab-6}
\begin{tabular}{|c|c|c|c|c|}
\hline
 force & orb & spin & total & R \\
\hline
 SVbas &  1.89 &  0.52 & 3.57 & 1.48 \\
\hline
SLy6 & 2.31  & 0.38 & 5.59 & 2.08 \\
\hline
SG2 & 2.15  & 0.48 &  3.74 & 1.42 \\
\hline
exper. \cite{Tamkas_NPA19} & & & 3.08 & \\
\hline
exper. \cite{Pitz_exp,Enders_PRC05} & & & 2.73(56) & \\
\hline
exper. $^{158}$Gd  \cite{Pitz_exp,Enders_PRC05} & & & 3.71(59) & \\
\hline
\end{tabular}
\\
\end{table}

 Table~\ref{tab-6} shows that, for SVbas and SG2, the summed M1 strength at OSR
energy region 2.5-4 MeV somewhat exceeds the experimental values for $^{156}$Gd
\cite{Tamkas_NPA19,Pitz_exp,Enders_PRC05}
but agrees with the measurements for the neighbor nucleus $^{158}$Gd
\cite{Tamkas_NPA19,Pitz_exp,Enders_PRC05}. Perhaps, a better agreement can be obtained after
incorporation of CCC. Anyway, already at QRPA level,  the agreement is quite acceptable.

Following Fig.~\ref{fig6_M1} and Table~\ref{tab-5}, the calculated
spin-flip mode in the OSR energy range is rather weak, and  so  $M1$ strength here is mainly
orbital. Nevertheless, a small spin fraction  is
important for the estimation of the total $M1$ strength. Fig.~\ref{fig6_M1} and
Table~\ref{tab-6} show  a large constructive interference of the orbital and spin fractions
($B(M1)_{\rm tot}>B(M1)_{\rm orb}+B(M1)_{\rm spin}$). The interference can be estimated by the factor
\begin{equation}\label{Rfac}
  R=\frac{B(M1)_{\rm tot}}{B(M1)_{\rm orb}+B(M1)_{\rm spin}}
\end{equation}
where $R=1, >1, <1$ for zero, constructive  and destructive interference. Table ~\ref{tab-6} shows that,
due to the interference, the total $M1$ strength at 2.5-4 MeV becomes 1.5-2 times larger than
the pure orbital strength.

Some time ago so called $M1$ "spin scissors" mode (located just below
the OSR) was predicted in deformed nuclei within macroscopic Wigner function moments approach
\cite{Bal_PRC18,Bal_PAN20}. The microscopic Skyrme QRPA analysis for $^{160,162,164}$Dy
and $^{232}$Th \cite{Nest_PRC_SSR,Nest_PAN_SSR} has shown that, indeed, some
low-energy spin-flip states can exist in some nuclei. However these states are
usually not collective and their current fields deviate from  the macroscopic spin-scissor flow.
As seen in Fig.~\ref{fig6_M1} and Table~\ref{tab-5}, such low-energy
spin-flip states can exist in deformed $^{156}$Gd as well. They lie at 2.21 MeV (SVbas), 2.46 MeV (SLy6) and
2.35 MeV (SG2). Each such state is represented by one 2qp configuration of spin-flip character.
SG2 calculations show that inclusion of $\textbf{J}^2$-term does not affect much these findings. 

\begin{table}
\centering
\caption{Calculated and experimental \cite{Tamkas_NPA19} summed B(M1) values
(in $\mu^2_N$) at 2.95 - 6.25 MeV in $^{156}$Gd.}
\label{tab-7}
\begin{tabular}{|c|c|c|c|}
\hline
 force & orb & spin & total  \\
\hline
SVbas & 4.21   &  2.72  &  6.89  \\
\hline
SLy6  & 4.50   & 1.35  & 7.22   \\
\hline
SG2 & 3.77  & 4.64  & 6.97  \\
\hline
exper. \cite{Tamkas_NPA19} & & & 3.12 \\
\hline
\end{tabular}
\end{table}

Fig.~\ref{fig6_M1} shows a significant discrepancy between calculated and experimental $M1$ strengths
 at 4-6 MeV. The experimental strength \cite{Tamkas_NPA19} at this energy
is almost absent while our calculations  give significant orbital, spin and total $B(M1)$-values.
Just for this reason, the calculated $M1$ strength summed at 2.95-6.25 MeV (6.9-7.2 $\mu^2_N$)
significantly overestimates the experimental value  3.12 $\mu^2_N$ (see  Table~\ref{tab-7}).
Note that a large $M1$ strength at 4-6 MeV was observed in $(e,e')$ reaction \cite{Richter90,Richter95}.
Moreover, a significant  spin-flip $M1$ strength at this energy range
in $^{156,158}$Gd  was reported in Skyrme QRPA studies \cite{Sarr96,Sasaki23}.
Our Skyrme QRPA calculations \cite{Nes_JPG10} and $(p,p')$ data \cite{Wor_exp94}
also give a significant spin-flip $M1$ strength  at 4-6 MeV in the neighbor nucleus $^{158}$Gd.
Following our Fig.~\ref{fig1}  and previous studies mentioned above, the interval 4-6 MeV
in $^{156,158}$Gd  {\it should include the  proton fraction of SFGR and thus embrace
an appreciable  $M1$ strength}. Inclusion of $\textbf{J}^2$-term in SG2 does not change this result.
In this connection, the absence of $M1$ strength at 4-6 MeV in
the NRF experiment for $^{156}$Gd \cite{Tamkas_NPA19} looks questionable.

\section{Conclusions}

$E1$ and $M1$ excitations in deformed  $^{156}$Gd were simultaneously investigated
within fully self-consistent QRPA calculations
with Skyrme forces SVbas, SLy6 and SG2. In the $E1$ channel, the
analysis covers low-lying $E1$ strength, called pygmy dipole resonance (PDR), giant dipole resonance (GDR),
and low-energy toroidal and compression $E1$ excitations. In $M1(K=1)$ channel, we consider
spin-flip giant resonance (SFGR), orbital scissors resonance (OSR), and low-energy spin-flip excitations.

The calculations well describe the energy and gross-structure (deformation splitting)
of $E1$ GDR. In PDR energy region, we reasonably reproduce the summed $E1$ strength at
2.95-6.25 MeV, measured in the NRF experiment
\cite{Tamkas_NPA19}. A dominance of $K=0$ strength over $K=1$ one
at this energy range points out the deformation splitting of these $K$-branches.
In agreement with our previous studies for deformed nuclei
\cite{Rep_EPJA17,Nest_PAN16,172Yb_E1tor,Ne_PRL18}, the present calculations confirm the
existence of the prominent toroidal $E1(K=1)$ resonance at PDR energy. Note that the vortical
toroidal mode can affect irrotational PDR E1 strength and its impact on astrophysical applications.
The compression low-energy E1 strength is weak and demonstrates a clear deformation splitting.

The calculations show a significant deformation-induced  admixture of E3 mode
(low-energy octupole resonance - LEOR  \cite{Har01,LEOR_exp,LEOR_MNS}) to dipole
$K=$0,1 states. This means that analysis of low-energy  E1 states in deformed nuclei
should take into account both dipole and octupole residual interaction, which automatically occurs
in self-consistent microscopic models. Note that the previous
systematic study  of electric dipole spectra in Gd isotopes within schematic QRPA \cite{Guliev20}
omits the octupole residual interaction.

The calculations reveal a strong constructive interference of the orbital and spin $M1$ modes
in OSR energy  region. This is important for estimation of the summed OSR M1 strength.

For $M1$ SFGR, the proton-neutron gross structure, impact of the residual interaction
and influence of central exchange $\textbf{J}^2$-term were analyzed. The dominance of
IV residual interaction over the IS one was justified. 
The calculations well reproduce proton-neutron splitting
of SFGR but overestimate energies of the proton and neutron $M1$ peaks.
Following our calculations with SG2, this shortcoming can be cured by taking into
account $\textbf{J}^2$-term. This result should be yet considered as preliminary since
it was obtained without i) refit of entire SG2 parameters and ii) inclusion of non-central
tensor forces. Anyway, it is shown that SFGR can be a valuable test for the impact of
$\textbf{J}^2$-term.

\begin{table*} 
\centering
\caption{Skyrme parameters for forces SVbas, SLy6 and SG2.}
\label{tab:Skyrme}       
\begin{tabular}{|c|c|c|c|c|c|c|c|c|c|c|}
\hline
Force & $t_0$ & $t_1$ & $t_2$ & $t_3$ & $t_4$ & $x_0$ & $x_1$ & $x_2$ & $x_3$ & $\alpha$ \\
\hline
SVbas & -1879 & 314 & 113 & 12527 & 125
      & 0.259 & -0.382 & -2.82 & 0.123 & 0.3 \\
 \hline
 SLy6 & -2479  & 462 & -449 & 13673  & 122
      & 0.825   & -0.465   & -1.0 & 1.355 & 0.167 \\
 \hline
SG2 & -2645 &   340 &  -42 &  15595 & 105
     &  0.09  &  -0.059 &  1.43 &  0.060 & 0.167 \\
\hline
\end{tabular}
\\
\end{table*}

Further, there was found a  puzzling discrepancy between calculated and experimental \cite{Tamkas_NPA19}
$M1$ strengths at 4-6 MeV. Following our calculations and previous theoretical \cite{Nes_JPG10,Sarr96,Sasaki23}
and experimental \cite{Richter90,Richter95,Wor_exp94} work, this energy interval should host the proton branch of SFGR and
so carry an impressive $M1$ strength. At the same time, the NRF experiment \cite{Tamkas_NPA19} delivers
almost no $M1$ strength at 4-6 MeV. The reason of the discrepancy is yet unclear.

\section*{ACKNOWLEDGEMENTS}
\label{s5}

We thank Dr. M. Tamkas for presentation of NRS experimental data.
J.K. appreciates the support by a grant of the Czech Science Agency, Project
No. 19-14048S. A. R. acknowledges support by the Slovak Research and Development
Agency under Contract No. APVV-20-0532 and by the Slovak grant agency VEGA
(Contract No. 2/0067/21).

\appendix
\numberwithin{equation}{section}
\section{Skyrme functional}
\label{Sec:_Skyrme}

The Skyrme functional has the form \cite{Ves_PRC09,Ben_RMP03,Stone_PPNP07}
\begin{eqnarray}
  \mathcal{H}_\mathrm{Sk}
  &=&
  \frac{b_0}{2} \rho^2- \frac{b'_0}{2} \sum\rho_{q}^2
  + \frac{b_3}{3} \rho^{\alpha+2}
  - \frac{b'_3}{3} \rho^{\alpha} \sum \rho^2_q
\nonumber
\\
 &&
 +b_1 (\rho \tau - \textbf{j}^{\;2})
 - b'_1 \sum(\rho_q \tau_q - \textbf{j}^{\;2}_q)
\nonumber
\\
 &&
 - \frac{b_2}{2} \rho\Delta \rho
 + \frac{b'_2}{2} \sum \rho_q \Delta \rho_q
\nonumber
\\
 &&
 - b_4 (\rho \nabla \cdot \textbf{J}\!+\!(\nabla\!\times\!\textbf{j}) \!\cdot\! \textbf{s})
\nonumber
\\
 &&
 - b'_4 \sum (\rho_q \nabla \cdot \textbf{J}_q\!
 +\!(\nabla\!\times\!\textbf{j}_q) \!\cdot\! \textbf{s}_q)
\nonumber
\\
  &&
  + \frac{\tilde{b}_0}{2} \textbf{s}^{\;2}
  - \frac{\tilde{b}'_0}{2} \sum \textbf{s}_{q}^{\;2}
+ \frac{\tilde{b}_3}{3} \rho^{\alpha} \textbf{s}^{\;2}
- \frac{\tilde{b}'_3}{3} \rho^{\alpha} \sum \textbf{s}^{\;2}_q
\nonumber
\\
  &&
 -\frac{\tilde{b}_2}{2} \textbf{s} \!\cdot\!
  \Delta \textbf{s} + \frac{\tilde{b}'_2}{2}
  \sum \textbf{s}_q \!\cdot\!\Delta \textbf{s}_q
\nonumber
\\
 &&
  +
  \tilde{b}_1
   (\textbf{s}\!\cdot\!\textbf{T}\!-\!\textbf{J}^{2})
  +
  \tilde{b}'_1
   \sum (\textbf{s}_q\!\cdot\!\textbf{T}_q
    \!-\!\textbf{J}_q^{2}) .
\label{eq:skyrme_funct}
\end{eqnarray}
Here $b_i$, $b'_i$, $\tilde{b}_i$, $\tilde{b}'_i$ are the force parameters.
The relation of these parameters with standard ones can be found in
Refs. \cite{Ves_PRC09,Ben_RMP03,Stone_PPNP07,Bonche_NPA87}.
Functional (\ref{eq:skyrme_funct}) includes time-even (nucleon
$\rho_q$, kinetic-energy $\tau_q$,
spin-orbit $\textbf{J}_q$) and time-odd (current $\textbf{j}_{ q}$, spin
$\textbf{s}_q$, spin kinetic-energy $\textbf{T}_q$) densities.
The label $q\in\{p,n\}$ $q$ denotes protons and neutrons.
Densities without index $q$, like $\rho = \rho_p +
\rho_n$, denote total densities. The contributions with $b_i$ (i=0,1,2,3,4) and
$b'_i$ (i=0,1,2,3) are the standard terms responsible for ground state
properties and electric excitations of even-even nuclei \cite{Ben_RMP03}.
The isovector spin-orbit interaction is usually linked to the
isoscalar one by $b'_4=b_4$. The terms with $\tilde{b}_i,\tilde{b}'_i$
($i$=0,1,2,3) represent the spin-isospin channel important for odd nuclei and magnetic
modes in even-even nuclei. The last line of (\ref{eq:skyrme_funct}) includes
 tensor spin-orbit terms $\propto\tilde{b}_1,\tilde{b}'_1$ which can affect both
 mean-field ground-state properties and magnetic modes. In general, these terms embrace
 contributions from both central-exchange interaction and non-central tensor interaction
 \cite{CS_PLB07,Les_PRC07,Bender_PRC09}. In our calculations, $\textbf{J}^2$
 is taken into account only for SG2 and only with central exchange contribution. 
In this case, $\tilde{b}_i$ and $\tilde{b}'_i$  are fully expressed through the standard 
Skyrme parameters  $t_1, t_2, x_1, x_2$ (see Eqs. (\ref{tb1})-(\ref{tb1'}) below)
and  have the values ${\tilde b}_1$=-40.1, ${\tilde b}'_1$=-47.8.

The parameters $b_i$, $b'_i$, $\tilde{b}_i$, $\tilde{b}'_i$ are related with standard Skyrme parameters
as \cite{Ves_PRC09,Ben_RMP03,Stone_PPNP07,Bonche_NPA87}.
\begin{eqnarray}
 b_0&=&t_0(1 + \frac{1}{2} x_0) ,
 \\
 b'_0&=&t_0(\frac{1}{2} + x_0) ,
 \\
 b_1&=&\frac{1}{4}[t_1(1 + \frac{1}{2}x_1) + t_2(1 + \frac{1}{2}x_2)] ,
 \\
 b'_1&=&\frac{1}{4}[t_1(\frac{1}{2} + x_1) - t_2(\frac{1}{2} + x_2)] ,
\\
 b_2&=&\frac{1}{8}[3t_1(1 + \frac{1}{2}x_1) - t_2(1 + \frac{1}{2}x_2)] ,
 \\
 b'_2&=&\frac{1}{8}[3t_1(\frac{1}{2} + x_1) + t_2(\frac{1}{2} + x_2)] ,
 \\
  b_3&=&\frac{1}{4}t_3(1 + \frac{1}{2} x_3) ,
 \\
 b'_3&=&\frac{1}{4}t_3(\frac{1}{2} + x_3) ,
 \\
  b_4&=&b'_4=\frac{1}{2}t_4 ,
\\
{\tilde b}_0&=&\frac{1}{2} t_0x_0 ,
\\
{\tilde b}'_0  &=&\frac{1}{2} t_0
\\
{\tilde b}_1&=&\frac{1}{8}(t_1x_1 + t_2x_2) ,
\label{tb1}
 \\
 \tilde{b}'_1&=&-\frac{1}{8}(t_1 - t_2) ,
 \label{tb1'}
\\
{\tilde b}_2 &=&\frac{1}{16}(3t_1x_1 - t_2x_2) ,
 \\
 \tilde{b}'_2&=&\frac{1}{16}(3t_1 + t_2) ,
 \\
{\tilde b}_3 &=&\frac{1}{8} t_3x_3 ,
 \\
 {\tilde b}'_3&=&\frac{1}{8} t_3 .
\end{eqnarray}

As seen from Table~\ref{tab:Skyrme}, Skyrme parameters $t_1, t_2, x_1, x_2$, entering
$\textbf{J}^2$ parameters ${\tilde b}_1$ and ${\tilde b}'_1$, are rather different
for SVbas, SLy6 and SG2.

\end{document}